\documentclass[conference,letterpaper]{IEEEtran}
\IEEEoverridecommandlockouts 

\usepackage{cite}
\usepackage[pdftex]{graphicx}   
\graphicspath{{./figs/}}
\usepackage[cmex10]{amsmath}
\usepackage{array}
\usepackage{amsfonts, amssymb}
\usepackage{amscd}
\usepackage{fixltx2e} 
\usepackage{url}

\newtheorem{theo}{Theorem}

\newtheorem{ex}{Example}

\newcommand{\Z}{{\mathbb{Z}}}
\newcommand{\zz}{\Z_2\Z_4}

\newcommand{\dd}{\displaystyle}

\newcommand{\codi}{{\cal C}}

\newcommand{\bx}{\mathbf{x}}
\newcommand{\bs}{\mathbf{s}}

\newcommand{\bh}{\mathbf{h}}

\newcommand{\bw}{\mathbf{w}}

\newcommand{\by}{\mathbf{y}}

\newcommand{\additive}{\Z_2\Z_4}

\newcommand{\dos}{{\mathbf{2}}}

\begin{document}

\title{Perfect $\additive$-linear codes in Steganography\thanks{This work was partially supported by the Spanish MICINN Grants MTM2009-08435, PCI2006-A7-0616, TSI2007-65406-C03-03 E-AEGIS, CONSOLIDER CSD2007-00004 ARES, and also by the \textit{Comissionat per a Universitats i Recerca del DIUE de la Generalitat de Catalunya} and the \textit{European Social Fund} with Grants 2009SGR1224 and FI-DGR.}}

\author{\IEEEauthorblockN{Helena Rif\`{a}\IEEEauthorrefmark{1},
Josep Rif\`{a}\IEEEauthorrefmark{2} and
Lorena Ronquillo\IEEEauthorrefmark{2}}
\IEEEauthorblockA{\IEEEauthorrefmark{1}Department of Computer Science and Multimedia,\\
Universitat Oberta de Catalunya,\\
Rb. del Poble Nou, 156, 08018-Barcelona, Spain.\\
Email: hrifa@uoc.edu}
\IEEEauthorblockA{\IEEEauthorrefmark{2}Department of Information and Communications Engineering,\\
Universitat Aut\`{o}noma de Barcelona,\\
08193-Cerdanyola del Vall\`{e}s, Spain.\\
Email: Josep.Rifa@autonoma.edu, Lorena.Ronquillo@autonoma.edu}
}

\maketitle

\begin{abstract}
Steganography is an information hiding application which aims to hide secret data imperceptibly into a commonly used media. Unfortunately, the theoretical hiding asymptotical capacity of steganographic systems is not attained by algorithms developed so far. In this paper, we describe a novel coding method based on $\zz$-linear codes that conforms to $\pm 1$-steganography, that is secret data is embedded into a cover message by distorting each symbol by one unit at most. This method solves some problems encountered by the most efficient methods known today, based on ternary Hamming codes. Finally, the performance of this new technique is compared with that of the mentioned methods and with the well-known theoretical upper bound.
\end{abstract}


\section{Introduction and preliminary results}\label{sec:intro}

{\it Steganography} is a scientific discipline within the field known as {\it data hiding}, concerned with hiding information into a commonly used media, in such a way that no one apart from the sender and the intended recipient can detect the presence of embedded data. A comprehensive overview of the core principles and the mathematical methods that can be used for data hiding can be found in \cite{moulin}.

An interesting steganographic method is known as {\it matrix encoding}, introduced by Crandall~\cite{cra} and analyzed by Bierbrauer et al.~\cite{bier}. Matrix encoding requires the sender and the recipient to agree in advance on a parity check matrix $H$, and the secret message is then extracted by the recipient as the syndrome (with respect to $H$) of the received cover object. This method was made popular by Westfeld~\cite{west}, who incorporated a specific implementation using Hamming codes in his F5 algorithm, which can embed $t$ bits of message in $2^t - 1$ cover symbols by changing, at most, one of them.

\medskip
There are two parameters which help to evaluate the performance of a
steganographic method over a cover message of $N$ symbols: the \textit{average distortion} $D=\frac{R_a}{N}$, where $R_a$
is the expected number of changes over uniformly distributed messages; and the
\textit{embedding rate} $E= \frac{t}{N}$, which is the amount of bits that can be
hidden in a cover message. In general, for the same embedding rate a method is better when the
average distortion is smaller. Following the terminology used by Fridrich et al.~\cite{fri}, the pair $(D,E)$ will be called {\it $CI$-rate}.

Furthermore, as Willems et al. in~\cite{WiMa}, we will also assume that a discrete source produces a sequence $\bx=(x_1,\ldots,x_N)$, where $N$ is the block length, each $x_i \in \aleph =\{0,1,\ldots,2^B-1\}$, and $B\in \{8,12,16\}$ depends on whether the source is a grayscale digital image, or a CD audio, etc.  The message $\bs \in \{1,\ldots,M\}$ we want to hide into a host sequence $\bx$ produces a composite sequence $\by = f(\bx,\bs)$, where $\by =(y_1,\ldots y_N)$ and each $y_i\in \aleph$. The composite sequence $\by$ is obtained from distorting $\bx$, and the distortion will be assumed to be a squared-error distortion (see~\cite{WiMa}). In these conditions, if information is only carried by the least significant bit (LSB) of each $x_i$, the appropriate solution comes from using binary Hamming codes~\cite{west}, improved using product Hamming codes~\cite{riri}. For larger magnitude of changes,
 but limited to $1$, that is, $y_i=x_i +c$, where $c\in \{0,+1,-1\}$, the situation is called ``$\pm 1$-steganography", and the information is carried by the two least significant bits. It is known that the embedding becomes statistically detectable
rather quickly with the increasing amplitude of embedding
changes. Therefore, our interest goes to avoid changes of amplitude greater than one. With this assumption, our steganographic scheme will be compared with the upper bound from~\cite{WiMa} for the embedding rate in ``$\pm 1$-steganography", given by $H(D)+D$, where $H(D)$ is the binary entropy function $H(D)= -D\log_2(D) - (1-D)\log_2(1-D)$ and $0\leq D\leq 2/3$ is the average distortion. A main purpose of steganography is designing schemes in order to approach this upper bound.

In most of the previous papers, ``$\pm 1$-steganography" has involved a ternary coding problem. Willems et al.~\cite{WiMa} proposed a schemed based on ternary Hamming and Golay codes, which were proved to be optimal. Fridrich and Lison\v{e}k~\cite{fri} proposed a method based on rainbow colouring graphs which, for some values, outperformed the scheme obtained by direct sum of ternary Hamming codes with the same average distortion.
However, both methods from~\cite{WiMa} and~\cite{fri} show a problem when dealing with extreme grayscale values, since they suggest making a change of magnitude greater than one in order to avoid having to apply the change $x_i-1$ and $x_i+1$ to a host sequence of value $x_i=0$ and $x_i=2^B-1$, respectively. Note that the kind of change they propose would obviously introduce larger distortion and therefore make the embedding more statistically detectable.

In this paper we also consider the $\pm 1$-steganography. Our new method is based on perfect $\additive$-linear codes which, although they are not linear, they have a representation using a parity check matrix that makes them as efficient as the Hamming codes. As we will show, this new method not only performs better than the one obtained by direct sum of ternary Hamming codes from~\cite{WiMa}, but it also deals better with the extreme grayscale values, because the magnitude of embedding changes is under no circumstances greater than one.

To make this paper self-contained, we review in Section~\ref{sec:additivecodes} a few elementary concepts on perfect $\additive$-linear codes, relevant for our study. The new steganographic method is presented in Section~\ref{sec:stegoz2z4}, whereas an improvement to better deal with the extreme grayscale values problem is given in Section~\ref{sec:anomalies}. Finally, the paper is concluded in Section~\ref{sec:conclusions}.

\section{Perfect $\additive$-linear codes}\label{sec:additivecodes}

In general, any non-empty subgroup $\codi$ of $\Z_2^\alpha \times \Z_4^\beta$ is a {\it $\zz$-additive code}, where $\Z_2^\alpha$ denotes the set of all binary
vectors of length $\alpha$ and $\Z_4^\beta$ is the set of all $\beta$-tuples in $\Z_4$. Let $C=\Phi(\codi)$, where
$\Phi: \Z_2^{\alpha}\times\Z_4^{\beta} \longrightarrow \Z_2^{n}$ is given by the map $$\Phi(u_1, \ldots, u_{\alpha} | v_1, \ldots, v_{\beta})=(u_1, \ldots, u_{\alpha} | \phi(v_1), \ldots, \phi(v_{\beta})),$$ where $\phi(0)=(0,0)$, $\phi(1)=(0,1)$, $\phi(2)=(1,1)$, and $\phi(3)=(1,0)$ is the usual Gray map from $\Z_4$ onto $\Z_2^2$.

A $\zz$-additive code $\codi$ is also isomorphic to an abelian structure like
$\Z_2^{\gamma}\times \Z_4^{\delta}$. Therefore, $\codi$ has $|\codi|=2^\gamma 4^\delta $ codewords, where $2^{\gamma+\delta}$ of them are of order two. We call such code $\codi$ a
{\it $\zz$-additive code of type $(\alpha,\beta;\gamma,\delta)$} and its
binary image $C$ is a {\it $\zz$-linear code of type
$(\alpha,\beta;\gamma,\delta)$}. Note that the Lee distance of a $\zz$-additive code $\codi$ coincides with the Hamming distance of the $\zz$-linear code $C=\Phi(\codi)$, and that the binary code $C$ does not have to be linear.

The {\it $\Z_2\Z_4$-additive dual code} of $\codi$, denoted by ${\cal
C}^\perp$, is defined as the set of vectors in $\Z_2^\alpha \times \Z_4^\beta$ that are orthogonal to every codeword in $\codi$, being the definition of inner product in $\Z_2^{\alpha}\times \Z_4^{\beta}$ the following:
\begin{equation} \label{inner}
  \langle u,v \rangle=2(\sum_{i=1}^{\alpha}
  u_iv_i)+\sum_{j=\alpha+1}^{\alpha+\beta}
u_jv_j\in \Z_4, \end{equation}
 where $u,v\in \Z_2^{\alpha}\times \Z_4^{\beta}$ and computations are made considering the zeros and ones in the $\alpha$ binary coordinates as quaternary zeros and ones, respectively.

The binary code $C_\perp=\Phi({\cal C}^\perp)$, of length $n=\alpha+2\beta$, is called the {\it
$\Z_2\Z_4$-dual code} of $C$.

A $\zz$-additive code $\codi$ is said to be {\it perfect} if code $C=\Phi(\codi)$ is a perfect $\zz$-linear code, that is all vectors in $\Z_2^{n}$ are within distance one from a codeword and the distance between two codewords is, at least, three.

For any $m\geq 2$ and each $\delta$ $\in$ $\{0,\ldots,\lfloor \frac{m}{2}\rfloor \}$ there exists a perfect $\zz$-linear code $C$ of binary length $n=2^m-1$, such that its $\zz$-dual code is of type $(\alpha,\beta;\gamma,\delta)$, where $\alpha=2^{m-\delta}-1$, $\beta=2^{m-1}-2^{m-\delta-1}$ and $\gamma=m-2\delta$ (note that the binary length can be computed as $n=\alpha+2\beta$). The above result is due to~\cite{br} and it allows us to write the parity check matrix $H$ of any $\zz$-additive perfect code for a given value of $\delta$. Matrix $H$ can be represented taking all possible vectors in $\Z_2^\gamma \times \Z_4^\delta$, up to sign changes, as columns. In this representation, there are $\alpha$ columns which correspond to the binary part of vectors in $\codi$, and $\beta$ columns of order four which correspond to the quaternary part. We agree on a representation of the $\alpha$ binary coordinates as coordinates in $\{0,2\} \in \Z_4$.

\section{Steganography based on perfect $\Z_2\Z_4$-linear codes}\label{sec:stegoz2z4}

Take a perfect $\zz$-linear code and consider its $\zz$-dual, which is of type $(\alpha,\beta;\gamma,\delta)$. As stated in the previous section, this gives us a parity check matrix $H$ which has $\gamma$ rows of order two and $\delta$ rows of order four.

For instance, for $m=4$ and according to~\cite{br}, there are three different $\zz$-additive perfect codes of binary length $n=2^4-1=15$ which correspond to the possible values of $\delta\in \{0,\ldots,\lfloor \frac{m}{2}\rfloor \} = \{0,1,2\}$. For $\delta=0$, the corresponding $\zz$-additive perfect code is the usual binary Hamming code, while for $\delta=2$ the $\zz$-additive perfect code has parameters $\alpha=3$, $\beta=6$, $\gamma=0$, $\delta=2$ and the following parity check matrix:
\begin{equation}\label{pcm}
H=\left ( \begin{array}{ccc|cccccc}
        2&0&2 & 0&1&1&1&1&2\\
	2&2&0 & 1&0&1&2&3&1
    \end{array}\right ).
\end{equation}

Let $\bh_i$, for $i \in \{1,\ldots,\alpha+\beta \}$, denote the $i$-th column vector of $H$.
Note that the all twos vector $\dos$ is always one of the columns in $H$ and, for the sake of simplicity, it will be written as the column $\bh_1$. We group the remaining first $\alpha$ columns in $H$ in such a way that, for any $2\leq i \leq (\alpha+1)/2$, the column vector $\bh_{2i}$ is paired up with its complementary column vector $\bar{\bh}_{2i} = \bh_{2i+1}$, where $\bar{\bh}_{2i}=\bh_{2i}+\dos$.

\medskip

To use these perfect $\zz$-additive codes in steganography take $N=2^{m-1}=\frac{\alpha+1}{2}+\beta$ and let $x=(x_1,\ldots,x_N)$ be an $N$-length source of grayscale symbols such that $x_i \in \aleph=\{0,1,\ldots,2^B-1\}$, where, for instance, $B=8$ for grayscale images.
We assume that a grayscale symbol $x_i$ is represented as a binary vector $(v_{7i},\ldots,v_{1i},v_{0i})$ such that

\begin{equation}\label{representation}
 x_i=\sum_{j=0}^{B/2-1}\phi^{-1}(v_{(2j+1)i} \;,\; v_{(2j)i}) \cdot 4^{j},
 \end{equation}

 where $\phi^{-1}()$ is the inverse of Gray map. We will use the two least significant bits (LSBs), $v_{1i},v_{0i}$, of every grayscale symbol $x_i$ in the source, for $i > 1$, as well as the least significant bit $v_{01}$ of symbol $x_1$ to embed the secret message.

\medskip

Each symbol $x_i$ will be associated with one or more column vectors $\bh_i$ in $H$, depending on the grayscale symbol:

\begin{enumerate}

\item Grayscale symbol $x_1$ is associated with column vector $\bh_1$ by taking the least significant bit $v_{01}$ of $x_1$.

\item Grayscale symbol $x_i$, for $2\leq i \leq (\alpha+1)/2$, is associated with the two column vectors $\bh_i$ and $\bar{\bh}_i$, by taking, respectively, the two least significant bits, $v_{1i}, v_{0i}$, of $x_i$.

\item Grayscale symbol $x_j$, for $\alpha < j \leq N$, is associated with column vector $\bh_{j+(\alpha-1)/2}$ by taking its two least significant bits $v_{1j},v_{0j}$  and interpreting them as an integer number $\phi^{-1}(v_{1j},v_{0j})$ in $\Z_4$.
\end{enumerate}

In this way, the given $N$-length packet $x$ of symbols is translated into a vector $\bw$ of $\alpha$ binary coordinates and $\beta$ quaternary coordinates.

The embedding process we are proposing is based on the matrix encoding method~\cite{cra,west}. The secret message can be any vector $\bs$ $\in$ $\Z_2^{\gamma}\times \Z_4^{\delta}$. Vector $\epsilon \cdot \bh_i$ indicates the changes needed to embed $\bs$ within $x$; that is $H{\bw}^T + \epsilon \cdot \bh_i=\bs$, where $\epsilon$ is an integer whose value will be described bellow, $H{\bw}^T$ is the syndrome vector of $\bw$ and $\bh_i$ is a column vector in $H$. The following situations can occur, depending on which column $\bh_i$ needs to be modified:

\begin{enumerate}
  \item If $\bh_i=\bh_1$, then the embedder is required to change the least significant bit of $x_1$ by adding or substracting one unit to/from $x_1$, depending on which operation will flip its least significant bit, $v_{01}$. 

 \item If $\bh_i$ is among the first $\alpha$ column vectors in $H$ and $2 \leq i \leq \alpha$, then $\epsilon$ can only be $\epsilon=1$. In this case, since $\bh_i$ was paired up with its complementary column vector $\bar{\bh}_i$, then this situation is equivalent to make $(v_{1i},1+v_{0i})$ or $(1+v_{1i},v_{0i})$, where $v_{1i}$ and $v_{0i}$ are the least significant bits of the symbol $x_i$ which had been associated with those two column vectors. Hence, after the inverse of Gray map, by changing one or another least significant bit we are actually adding or subtracting one unit to/from $x_i$. Note that a problem may crop up at this point when we need to add $1$ to a symbol $x_i$ of value $2^B-1$ or, likewise, when $x_i$ has value $0$ and we need to subtract $1$ from it.

\item If $\bh_i$ is one of the last $\beta$ columns in $H$ we can see that this situation corresponds to add $\epsilon \in \{0,1,2,3\}$ to $x_{i-(\alpha-1)/2}$. Note that because we are using a $\zz$-additive perfect code, $\epsilon$ will never be $2$. Hence, the embedder should add ($\epsilon=1$) or subtract ($\epsilon=3$) one unit to/from symbol $x_{i-(\alpha-1)/2}$. Once again, a problem may arise with the extreme grayscale values.

\end{enumerate}

\bigskip

\begin{ex} \label{ex:noAnomalies} Let $x=(239,251,90,224,226,187,229,180)$ be an $N$-length source of grayscale symbols, where $x_i \in \{0,\ldots,255\}$ and $N=8$, and let $H$ be the matrix in (\ref{pcm}). The source $x$ is then translated into the vector $\bw=(010|202310)$ in the way specified above. Let $\bs={0 \choose 2}$ be the vector representing the secret message we want to embed in $x$. We then compute $H{\bw}^T={2 \choose 3}$ and see, by the matrix encoding method, that $\epsilon=3$ and $\bh_i=\bh_9$. According to the method just described, we should apply the change $x_8-1$. In this way, $x_8$ becomes $x_8=179$, and then $\bw=(010|202313)$, which has the expected syndrome ${0 \choose 2}$.
\end{ex}

\medskip

As already mentioned at the beginning of this paper, the problematic cases related to the extreme grayscale values are also present in the methods from \cite{fri} and \cite{WiMa}, but their authors assume that the probability of gray value saturation is not too large. We argue that, though rare, this gray saturation can still occur. However, in order to compare our proposal with these others we will not consider these problems either until the next section.
Therefore, we proceed to compute the values of the average distortion $D$ and the embedding rate $E$.

Our method is able to hide any secret vector $\bs$ $\in$ $\Z_2^{\gamma}\times \Z_4^{\delta}$ into the given $N$ symbols. Hence, the embedding rate is $(\gamma+2\delta)$ bits per $N$ symbols, $E=\dd \frac{\gamma+2\delta}{N}=\frac{m}{2^{m-1}}$.

Concerning the average distortion $D$, we are using a perfect code of binary length $2^{m}-1$, which corresponds to $N=2^{m-1}$ grayscale symbols. There are $N-1$ symbols $x_i$, for $2 \leq i \leq N$,  with a probability $2/2^{m}$ of being subjected to a change; a symbol $x_1$ with a probability $1/2^{m}$ of being the one changed; and, finally, there is a probability of $1/2^{m}$ that neither of the symbols will need to be changed to embed $\bs$. Hence, $D=\dd \frac{2N-1}{N2^{m}}=\frac{2^m-1}{2^{2m-1}}$.

The described method has a $CI$-rate $(D_m,E_m) = \left ( \dd \frac{2N-1}{2N^2}, \frac{1+\log(N)}{N} \right )$, where $N=2^{m-1}$ and $m$ is any integer $m\geq 2$.
We are able to generate a specific embedding scheme for any value of $m$ but not for any $CI$-rate.

\begin{figure}[htp]
\centering
\includegraphics[scale=0.4]{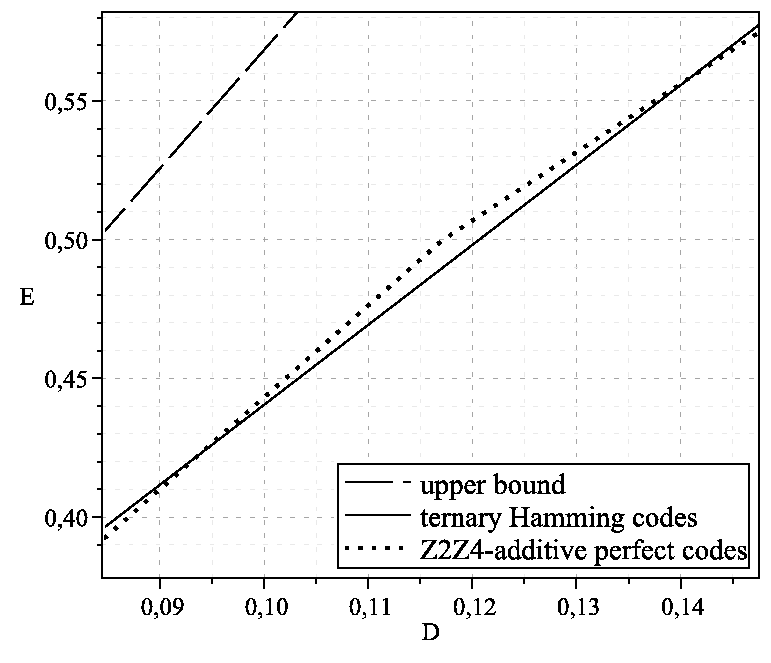}
\caption{$CI$-rate $(D,E)$, for $B=8$, of steganographic methods based on ternary Hamming codes and on $\zz$-additive perfect codes.}\label{fig:graphicWoAnom}
\end{figure}

With the aim of improving this situation, convex combinations of $CI$-rates of two codes related to their direct sum are extensively treated in~\cite{fri}. Actually, it is possible to choose the $D$ coordinate and cover more $CI$-rates by taking convex combinations. Therefore, if $D$ is a non-allowable parameter for the average distortion we can still take $D_1 < D < D_2$, where $D_1,D_2$ are two contiguous allowable parameters, and by means of the direct sum of the two codes with embedding rate $E_1$ and $E_2$, respectively, we can obtain a new $CI$-rate $(D,E)$, with $D=\lambda D_1+(1-\lambda) D_2$ and $E=\lambda E_1+(1-\lambda) E_2$. From~a graphic point of view, this is equivalent to draw a line between two contiguous points $(D_1,E_1)$ and $(D_2,E_2)$, as it is shown in \figurename~\ref{fig:graphicWoAnom}.

In the following theorem we claim that the $CI$-rate of our method improves the one given by direct sum of ternary Hamming codes from~\cite{WiMa}.

\begin{theo}
For $m\geq 4$, the $CI$-rate given by the method based on $\zz$-additive perfect codes improves the $CI$-rate obtained by direct sum of ternary Hamming codes with the same average distortion.
\end{theo}
\begin{IEEEproof}
Optimal embedding (of course, in the allowable values of $D$) can be obtained by using ternary codes, as it is shown in~\cite{WiMa}. The $CI$-rate of these codes is
$(D_\mu,E_\mu) = \left ( \dd \frac{2}{3^\mu}, \frac{2\mu}{3^\mu-1} \right )$
for any integer $\mu$. Our method, based on $\zz$-additive perfect codes, has $CI$-rate $(D_m,E_m) = \left ( \dd \frac{2N-1}{2N^2}, \frac{1+\log(N)}{N} \right )$, for any integer $m\geq 2$ and $N=2^{m-1}$.

Take, for any $m\geq 2$, two contiguous values for $\mu$ such that $D_{\mu+1} < D_m < D_{\mu}$ and write $D_m=\lambda D_{\mu+1}+(1-\lambda) D_\mu$, where $0\leq \lambda \leq 1$.

We want to prove that, for $m\geq 4$, we have  $E_m \geq \lambda E_{\mu+1}+(1-\lambda) E_\mu$, which is straightforward. However, since it is neither short nor contributes to the well understanding of the method, we do not include all computations here. The graphic bellow compares the $CI$-rate of the method based on ternary Hamming codes with that one based on $\zz$-additive perfect codes. As one may see in this graphic, for some values of the average distortion $D$, the scheme based on $\zz$-additive perfect codes has greater embedding rate $E$ than the one based on ternary Hamming codes.
\end{IEEEproof}

\textbf{Remark:} The same argumentation can be used and the same conclusion can be reached taking $q$ instead of $3$ and comparing our method with the method described in \cite{fri}.

\section{Solving the extreme grayscale values problem}\label{sec:anomalies}

In Section~\ref{sec:stegoz2z4} we described a problem which may raise when, according to our method, the embedder is required to add one unit to a source symbol $x_i$ containing the maximum allowed value ($2^B-1$), or to substract one unit from a symbol $x_i$ containing the minimum allowed value, $0$. To face this problem, we will use the complementary column vector $\bar{\bh}_i$ of columns $\bh_i$ in matrix $H$, where $\bar{\bh}_i = 3\bh_i + \textbf{2}$ and $\bh_i$ is among the last $\beta$ columns in $H$. Note that $\bh_i$ and $\bar{\bh}_i$ can coincide.

The first $\alpha$ column vectors in $H$ will be paired up as before, and the association between each $x_i$ and each column vector $\bh_i$ in $H$ will be also the same as in Section~\ref{sec:stegoz2z4}. However, given an $N$-length source of grayscale symbols
$x=(x_1,\ldots,x_N)$, a secret message $\bs \in \Z_2^{\gamma}\times \Z_4^{\delta}$ and the vector $\epsilon \cdot \bh_i$, such that $H{\bw}^T + \epsilon \cdot \bh_i= \bs$, indicating the changes needed to embed $\bs$ within $x$, we can now make some variations on the kinds of changes to be done for the specific problematic cases:

\begin{itemize}

\item If $\bh_i$ is among the first $\alpha$ columns in $H$, for $2 \leq i \leq \alpha$, and the embedder is required to add $1$ to a symbol $x_i=2^B-1$, then the embedder should instead substract $1$ from $x_i$ as well as perform the appropiate operation ($+1$ or $-1$) over $x_1$ to have $v_{01}$ flipped. Likewise, if the embedder is required to substract $1$ from a symbol $x_i=0$, then (s)he should instead add $1$ to $x_i$ and also change $x_1$ to flip $v_{01}$.

\item If $\bh_i$ is one of the last $\beta$ columns in $H$, and the embedder has to add $1$ to a symbol $x_i=2^B-1$, (s)he should instead substract $1$ from the grayscale symbol associated to $\bar{\bh}_i$ and also change $x_1$ to flip $v_{01}$.
If the method requires substracting $1$ from $x_i=0$, then we should instead add $1$ to the symbol associated to $\bar{\bh}_i$ and, again, change $x_1$ to flip $v_{01}$.
\end{itemize}

\bigskip

\begin{ex}\label{ex:Anomalies} Let $x=(239,251,90,224,226,187,229,0)$ be an $N$-length source of grayscale symbols, where $x_i \in \{0,\ldots,255\}$ and $N=8$, and let $H$ be the matrix (\ref{pcm}). As in Example~\ref{ex:noAnomalies}, the packet $x$ is translated into vector $\bw=(010|202310)$, and $\bs={0 \choose 2}$. However, note that now we are not able to make $x_8-1$ because $x_8=0$. Instead of this, we will add one unit to $x_3$, which is the symbol associated with $\bar{\bh}_9=\bh_4$, and substract one unit from $x_1$ so as to have its least significant bit flipped. Therefore, we obtain  $x=(238,251,91,224,226,187,229,0)$ and then $\bw=(110|302310)$, which has the desired syndrome.
\end{ex}

\medskip

The method above described has the same embedding rate $E=\dd \frac{m}{2^{m-1}}$ as the one from Section~\ref{sec:stegoz2z4} but a slightly worse average distortion. We will take into account the squared-error distortion defined in~\cite{WiMa} for our reasoning.

As before, among the total number of grayscale symbols $N=2^{m-1}$, there are $N-1$ symbols $x_i$, for $2 \leq i \leq N$, with a probability $2/2^{m}$ of being changed; a symbol $x_1$ with a probability $1/2^{m}$ of being the one changed; and, finally, there is a probability of $1/2^{m}$ that neither of the symbols will need to be changed.

As one may have noted in this scheme, performing a certain change to a symbol $x_i$, associated with a column $\bh_i$ in $H$, has the same effect as performing the opposite change to the grayscale symbol associated with $\bar{\bh}_i$ and also changing the least significant bit $v_{01}$ of $x_1$. This means that with probability $\frac{2^B-2}{2^B}$ we will change a symbol $x_i$, for $2 \leq i \leq N$, a magnitude of $1$; and with probability $\frac{2}{2^B}$ we will change two other symbols also a magnitude of $1$. Therefore, $R_a = \dd (N-1) \frac{2}{2^m} \left ( \frac{2^B-2}{2^B}+2\frac{2}{2^B} \right ) + \frac{1}{2^m}$ and the average distortion is thus $D= \dd \frac{ 2N - 1 + \frac{N-1}{2^{B-2}} }{ N2^m }$. Hence, the described method has $CI$-rate $\dd (D_m,E_m) = \left ( \frac{2N-1 + \frac{N-1}{2^{B-2}}}{2N^2} , \frac{1+\log(N)}{N} \right ) $.

\medskip

As we have already mentioned, the problem of grayscale symbols with $0$ and $2^B-1$ values was previously detected in both \cite{fri} and \cite{WiMa}. With the aim of providing a possible solution to this problem, the authors suggested to perform a change of a magnitude greater than $1$. However, the effects of doing this were are out of the scope of $\pm 1$-steganography.

In the remainder of this section we proceed to compare the $CI$-rate of our method with the $CI$-rate that those methods would have if their proposed solution was implemented.

\medskip

The scheme presented by Willems et al.~\cite{WiMa} is based on ternary Hamming codes, which are known to have length $n=(3^\mu-1)/2$, where $\mu$ denotes the number of parity check equations. Let us assume that whenever the embedder is required to perform a change ($+1$ or $-1$) that would lead the corresponding symbol $x_i$ to a non-allowed value, then a change of magnitude $2$ ($-2$ or $+2$) is made instead. While the embedding rate $E$ of this scheme would still be $E=\dd \frac{2\mu \log(3)}{3^\mu-1}$, the average distortion $D$ would no longer be $D=\frac{2}{3^\mu}$. The actual expected number of changes $R_a$ is computed by noting that a symbol will be changed with probability $\frac{3^\mu-1}{3^\mu}$, and will not with probability $\frac{1}{3^\mu}$. Among the cases in which a symbol would need to be changed, there is a probability of $\frac{2^B-2}{2^B}$ that a symbol will be changed a magnitude of $1$, and a probability of $\frac{2}{2^B}$ that it will be changed a magnitude of $2$. By the squared-error distortion, $R_a=\frac{3^\mu-1}{3^\mu} \left ( \frac{2^B-2}{2^B}\cdot1+\frac{2}{2^B}\cdot2^2 \right )$ and therefore $D=\dd \frac{2}{3^\mu} \left ( 1+\frac{3}{2^{B-1}} \right )$.

\medskip

Fridrich and Lison\v{e}k propose in their paper to pool the grayscale symbols source $x$ into cells of size $d$, then rainbow colour these cells and apply a $q$-ary Hamming code, where $q=2d+1$ is a prime power.
They measure the distortion by counting the maximum number of embedding changes, thus just considering the covering radius of the $q$-ary Hamming codes. However, we will now consider the average number of embedding changes (see~\cite{friEff}). As Willems et at., the authors from \cite{fri} also suggest to perform a change of magnitude $q-1>1$ to solve the extreme grayscale values problem. If this is done, the embedding rate would still be the same, $E=\dd \frac{2\mu \log(q)}{q^\mu-1}$, but the average distortion would now be $D=\frac{2}{q^{\mu}} \left ( \frac{2^B-2}{2^B} + \frac{2}{2^B} \cdot (q-1)^2 \right ) = \dd \frac{2}{q^{\mu}} \left ( 1 + \frac{q(q-2)}{2^{B-1}} \right )$.

\medskip

\begin{figure}[htb!]
\centering
\includegraphics[scale=0.4]{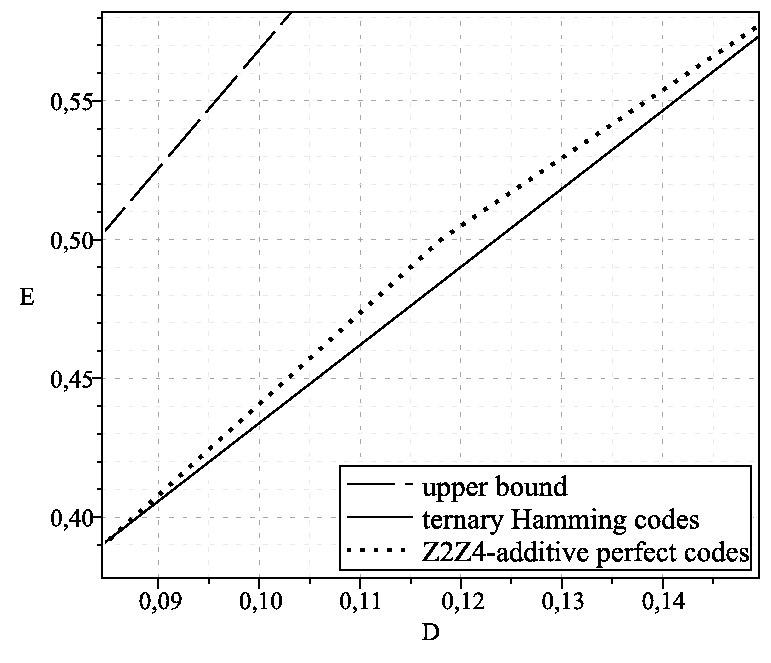}
\caption{$CI$-rates $(D,E)$, for $B=8$, of steganographic methods based on ternary Hamming codes and on $\zz$-additive perfect codes, when they are dealing with the extreme grayscale values problem described in Section~\ref{sec:anomalies}.}\label{fig:graphicWAnom}
\end{figure}

One can see in \figurename~\ref{fig:graphicWAnom} how our steganographic method for $\zz$-additive perfect codes deals with the extreme grayscale values problem, for some values of $D$, better than those using ternary Hamming codes ($q=3$) from \cite{fri} and \cite{WiMa}.

\section{Conclusions}\label{sec:conclusions}

In this paper, we have presented a new method for $\pm 1$-steganography, based on perfect $\additive$-linear codes. These codes are non-linear but still there exists a parity check matrix representation that makes them efficient to work with.

As we have shown in sections~\ref{sec:stegoz2z4} and~\ref{sec:anomalies}, this new scheme outperforms the one obtained by direct sum of ternary Hamming codes (see~\cite{WiMa}) as well as the one obtained after rainbow colouring graphs by using $q$-ary Hamming codes for $q=3$.

If we consider the special cases in which the technique might require to substract one unit from a grayscale symbol containing the minimum allowed value, or to add one unit to a symbol containing the maximum allowed value, our method performs even better than those aforementioned schemes. This is so because unlike them, our method never applies any change of magnitude greater than $1$, but two changes of magnitude $1$ instead, which is better in terms of distortion. Therefore, our method makes the embedding less statistically detectable.

As for further research, since the approach based on product Hamming codes in~\cite{riri} improved the performance of basic LSB steganography and the basic $F5$ algorithm, we would also expect a considerable improvement of the $CI$-rate by using product $\additive$-additive codes or subspaces of product $\additive$-additive codes in $\pm 1$-steganography.

\end{document}